\documentclass[]{spie}  

\usepackage{amsmath,amsfonts,amssymb}
\usepackage{graphicx}
\usepackage[colorlinks=true, allcolors=blue]{hyperref}
\usepackage{graphicx}
\usepackage{float}
\usepackage{pifont}
\title{Zonal-Aware Prostate Cancer Detection and Classification via Hadamard-Fused Mamba-UNet}

\author[a]{Emadeldeen Hamdan\textsuperscript{*}}
\author[b]{Gorkem Durak}
\author[b]{Muhammed Enes Tasci}
\author[b]{Halil Ertugrul Aktas}
\author[c]{Fergan Bol}
\author[b]{Elif Bilge Gok}
\author[d]{Mehmet Eralp Akpinar}
\author[e]{Ali Berke Vural}
\author[b]{Andrea M. Bejar}
\author[b]{Wanying Dou}
\author[b]{Elif Keles}
\author[f]{Frank H Miller}
\author[g]{Hiten D. Patel}
\author[g]{Ashley E Ross}
\author[f]{Hatice Savas}
\author[a]{Ahmet Enis Cetin}
\author[b]{Ulas Bagci}
\affil[a]{Electrical and Computer Engineering Department, University of Illinois Chicago, IL, USA}
\affil[b]{Machine and Hybrid Intelligence Lab, Northwestern University, IL, USA}
\affil[c]{Department of Radiology, University of Health Sciences Istanbul Bakirkoy Dr. Sadi Konuk Training and Research Hospital, Turkey}
\affil[d]{Gazi University Faculty of Medicine, Turkey}
\affil[e]{University of Ottawa, Canada}
\affil[f]{Department of Radiology, Northwestern University, IL, USA}
\affil[g]{Department of Urology, Feinberg School of Medicine, Northwestern University, IL, USA}

\authorinfo{*Corresponding author: Emadeldeen Hamdan, E-mail: ehamda3@uic.edu}

\begin{document} 
\maketitle

\begin{abstract}
Accurate clinically significant prostate cancer (csPCa) detection from prostate MRI requires a detailed understanding of the prostate anatomical context. Deep learning approaches have achieved promising results in zonal and lesion segmentation; however, the joint analysis of prostate zonal anatomy and lesion morphology in a unified framework remains relatively underexplored. In this work, we propose ZH-Mamba, a multi-stage zonal-aware csPCa detection framework consisting of two H-Mamba UNets models for prostate zonal segmentation and lesion localization, followed by an anatomy-aware Convolutional Neural Network classifier for csPCa detection. The proposed model leverages a learnable Hadamard Transform-based fusion mechanism to enhance zonal and lesion feature representations without increasing computational complexity. The learned fusion patterns consistently emphasize prostate regions, suggesting that the proposed fusion captures anatomically relevant feature representations. We demonstrate a strong zonal segmentation performance on the PI-CAI dataset, reaching Dice scores of \textbf{0.95}, \textbf{0.85}, and \textbf{0.95} for whole gland (WG), peripheral zone (PZ), and transition zone (TZ), respectively, providing robust anatomical priors for downstream csPCa detection. Experimental results on an independent internal dataset confirm the generalization capability of the proposed zonal segmentation model. For lesion localization, the H-Mamba UNet achieves a competitive Dice score of \textbf{0.70}. Integrating the zonal masks and lesion localization, the proposed ZH-Mamba framework achieves an AUC of \textbf{0.86}, Average Precision ($\mathrm{AP}_{\mathrm{Patient}}$) of \textbf{0.72}, and F1 score of \textbf{0.76}, demonstrating the effectiveness of zonal-aware csPCa detection.
\end{abstract}

\keywords{Prostate MRI, csPCa, Zonal segmentation, Lesion localization, Mamba, Hadamard feature fusion}

\section{INTRODUCTION}
\label{sec:intro}  

Lesion appearance and diagnostic assessment differ considerably across prostate anatomical regions, particularly between the peripheral zone (PZ) and transition zone (TZ), making zonal anatomy an important component of clinically significant prostate cancer (csPCa) diagnosis~\cite{lee2015biologic}. This distinction is clinically important because the peripheral and transition zones exhibit different tissue characteristics, cancer prevalence, and imaging appearances, providing complementary anatomical information for prostate cancer assessment~\cite{wood2026location,imai2024optimizing}. Multiparametric magnetic resonance imaging (mpMRI), consisting of T2-weighted imaging (T2WI), diffusion-weighted imaging (DWI), apparent diffusion coefficient (ADC) maps, and dynamic contrast-enhanced (DCE) imaging, is the standard modality for csPCa detection, while biparametric MRI (bpMRI), which relies only on T2WI and DWI/ADC, is increasingly adopted as a contrast-free alternative~\cite{kang2019abbreviated,thestrup2016biparametric}.

Recent deep learning approaches have demonstrated promising performance in prostate MRI analysis, including prostate zonal segmentation~\cite{wu2022automatic,hung2022cat}, lesion segmentation~\cite{zhang2025generalist,yan2025semi}, and clinically significant prostate cancer (csPCa) classification~\cite{hamdan2026hybrid}. Convolutional neural networks (CNNs)~\cite{isensee2024nnu}, Transformer-based architectures~\cite{hung2022cat,wang2025sam}, and Mamba-based state space models (SSMs)~\cite{du2024mixed,wang2024mamba} have achieved strong segmentation and detection performance on public datasets such as PI-CAI~\cite{saha2024artificial}. However, despite these advances, most existing approaches address zonal segmentation, lesion localization, and cancer classification as separate tasks, preventing anatomical zonal information from being propagated throughout the diagnostic pipeline~\cite{yuan2025z}.

\begin{figure*}[t]
\centering
\includegraphics[width=1\textwidth, height=5cm]{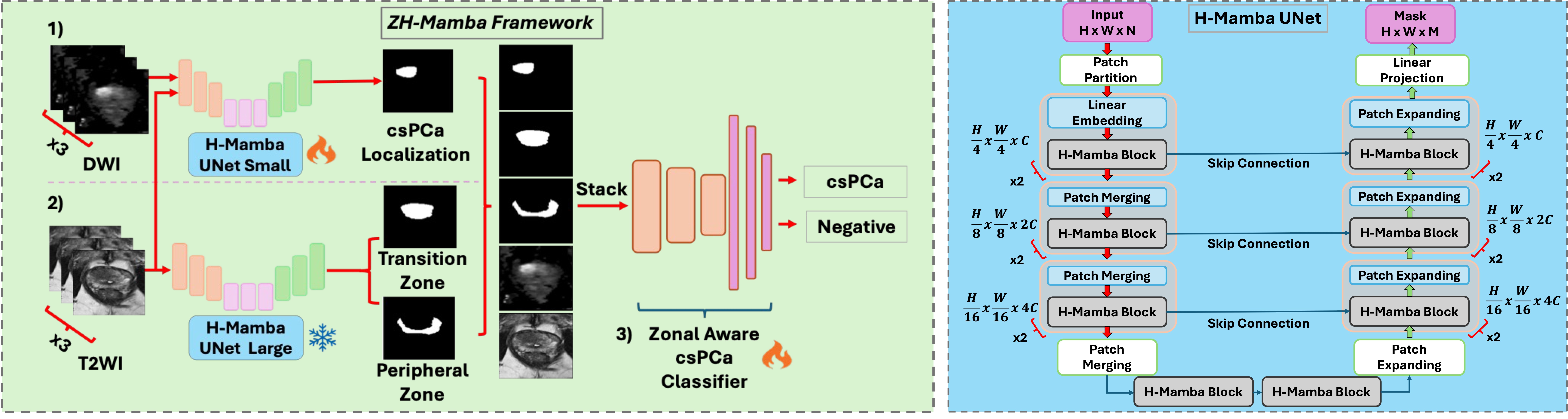}
\caption{\textbf{ZH-Mamba framework:} Overview of the proposed zonal-aware framework for csPCa detection. \textbf{H-Mamba UNet:} Integration of Hadamard-enhanced Mamba blocks, where learnable Hadamard features are fused with VSSM representations.}
\label{fig:model}
\end{figure*}

In this work, we propose ZH-Mamba, a multi-stage zonal-aware framework integrating prostate zonal segmentation, lesion localization, and csPCa classification. As shown in Figure.~\ref{fig:model}, two Hadamard-fused Mamba-UNets (H-Mamba UNets) perform zonal segmentation and lesion localization, followed by a CNN that integrates MRI, predicted lesion information, and zonal anatomy for csPCa classification. H-Mamba introduces a learnable fusion mechanism based on the two-dimensional Hadamard Transform~\cite{agaian2011hadamard}, providing global transform-domain feature interactions with minimal additional parameters. By propagating anatomical information through the diagnostic pipeline, ZH-Mamba achieves accurate zonal segmentation on PI-CAI and an independent internal dataset, strong lesion localization, and robust patient-level csPCa detection using only a small number of MRI slices.



Our main contributions of this work are three-folds. Firstly, we propose ZH-Mamba, a unified zonal-aware framework that integrates prostate zonal segmentation, lesion localization, and clinically significant prostate cancer (csPCa) classification within a multi-stage diagnostic pipeline. Secondly, we introduce a learnable Hadamard fusion mechanism for H-Mamba UNets that enhances feature representations through global transform-domain interactions with minimal additional parameters. Finally, we demonstrate that accurate prostate zonal segmentation, lesion localization, and robust patient-level csPCa detection can be achieved using only a small number of MRI slices.





\section{Methodology}
\noindent\textbf{Zonal Aware Hadamard-Mamba Overview.} 
Figure~\ref{fig:model} provides an overview of the proposed Zonal-Aware Hadamard-Fused Mamba (ZH-Mamba) framework that is designed to explicitly propagate anatomical zonal information throughout the prostate cancer diagnostic pipeline. ZH-Mamba consists of three stages: (1) a lightweight H-Mamba UNet for lesion localization, (2) a large H-Mamba UNet for prostate zonal segmentation, and (3) a CNN that combines the predicted zonal masks and lesion localization with the corresponding central T2WI and DWI images for clinically significant prostate cancer (csPCa) classification.

\noindent\textbf{Mamba SSM.} 
State Space Models (SSMs) provide a framework for modeling long-range dependencies through a latent state representation~\cite{gu2024mamba}. The continuous-time formulation is given by 
$h'(t) = A h(t) + B x(t), y(t) = C h(t) + D x(t)$,
where $h(t) \in \mathbb{R}^{d}$ denotes the latent state, and $A \in \mathbb{R}^{d\times d}$, $B \in \mathbb{R}^{d\times C}$, $C \in \mathbb{R}^{C\times d}$, and $D \in \mathbb{R}^{C\times C}$ are learnable parameters. 

For an input feature tensor $X \in \mathbb{R}^{N \times C}$, the state-space layer produces an output representation $Y = \mathrm{SSM}(X)$, enabling efficient long-range feature modeling with linear complexity $O(NC)$ compared to self-attention mechanisms with quadratic complexity $O(N^2)$. These SSM layers form the core building block of the proposed H-Mamba UNet.

\noindent\textbf{Hadamard Transform.} The Hadamard transform (HT) is an orthogonal transform that decomposes signals into a set of orthogonal binary basis functions~\cite{horadam2012hadamard} and have demonstrated effectiveness across various image analysis applications~\cite{hamdan2025real}. The HT has a computational complexity $O(N \log N)$. The normalized Hadamard matrix of order $2^m$ is defined recursively as
\begin{equation}
H_m = \frac{1}{\sqrt{2}}
\begin{bmatrix}
H_{m-1} & H_{m-1} \\
H_{m-1} & -H_{m-1}
\end{bmatrix}, 
\quad H_0 = [1],
\end{equation}
where $H_m H_m^{T} = I$.

In this work, the Walsh Hadamard transform (WHT)~\cite{ahmed1975walsh} is integrated into the H-Mamba-UNet as a mechanism for global feature mixing~\cite{walia2025fusion}. 

\noindent\textbf{H-Mamba UNets} To balance segmentation accuracy and computational efficiency, two task-specific H-Mamba UNets are employed. A larger network is dedicated to prostate zonal segmentation, while a lightweight network performs lesion localization.
Both H-Mamba UNets follow an encoder–decoder architecture with skip connections similar to conventional UNet models~\cite{wang2024mamba,zhu2026vhu,soulami2021breast} and operate on hierarchical patch representations through successive patch partitioning and merging operations.

The zonal segmentation network takes as input a stack of three consecutive T2-weighted slices $X_{T2} \in \mathbb{R}^{H \times W \times 3}$ and produces PZ and TZ segmentation masks: $M_{PZ}, M_{TZ} \in \mathbb{R}^{H \times W}$. Similarly, the lesion localization network takes as input three consecutive T2WI slices together with the corresponding three DWI slices, $X_{T2} \in \mathbb{R}^{H \times W \times 3}$ and $X_{DWI} \in \mathbb{R}^{H \times W \times 3}$, and predicts a lesion localization mask $M_{L} \in \mathbb{R}^{H \times W}$, as illustrated in Fig.~\ref{fig:model}. The choice of the number of slices used in each model is justified through the ablation study in Section~\ref{experiments}.

\begin{figure*}[t]
\centering
\includegraphics[width=0.8\textwidth, height=4cm]{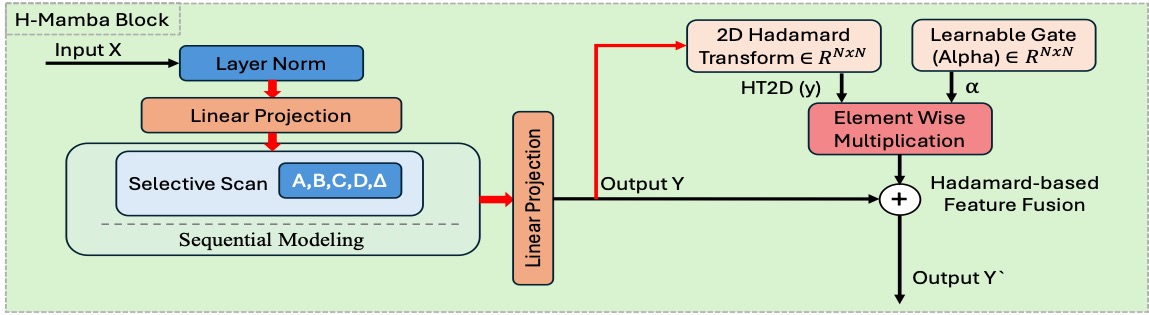}
\caption{\textbf{H-Mamba feature fusion.} The VSSM output is transformed using the 2D Walsh--Hadamard Transform (WHT) and modulated by a learnable gate $\alpha$ before residual fusion with the original spatial features.}
\label{fig:Fusion}
\end{figure*}

\noindent\textbf{Transform-Based Feature Mixing.}  Although the VSSM layer effectively captures long-range spatial dependencies, its sequential state-space modeling is less effective at representing high-frequency image components. In this work, we introduce a learnable Hadamard-based feature fusion mechanism that jointly exploits spatial and transform-domain representations~\cite{zhang2025spatial,zhu2024vision}. 

Let $Y \in \mathbb{R}^{N \times M \times C}$ denote the output feature tensor of the VSSM layer. The 2D Hadamard transform is applied independently to each channel. The feature tensor $Y$ of size $N$ by $M$, i.e, $Y \in R^{N \times M}$ is transformed using a two-dimensional Hadamard transform: $WHT(Y) = H_N \, Y \, H_M^{T},$ where $WHT(Y)$ is the Walsh Hadamard transform of $Y$, $H_N$ is the Hadamard matrix of size  $N \times N$, and $H_M$ is the Hadamard matrix of size  $M \times M$.

As illustrated in Figure.~\ref{fig:Fusion}, the transform-domain features are modulated by a learnable fusion coefficients Alpha: $\alpha \in \mathbb{R}^{H \times W \times C}$ through element-wise multiplication: $Z = \alpha \odot WHT(Y)$, where $\odot$ denotes element-wise multiplication. The fused feature representation is obtained through residual addition: $Y' = Y + Z = Y + \alpha \odot WHT(Y)$. The proposed fusion introduces only HWC additional learnable parameters while providing global transform-domain interactions.

\noindent\textbf{Zonal-Aware csPCa Classification.} Unlike conventional slice-level classifiers that rely solely on MRI appearance, the proposed classifier explicitly integrates anatomical zonal information (PZ and TZ), lesion localization, and MRI intensity information for csPCa prediction. The predicted zonal masks and lesion localization mask are concatenated with the MRI images to form: 
\begin{equation}
X_{cls} =
\left[
X_{T2}^{(c)}, \,
X_{DWI}, \,
M_{PZ}, \,
M_{TZ}, \,
M_{L}
\right]
\in \mathbb{R}^{H \times W \times 5}
\end{equation}
where $X_{T2}^{(c)}$ and $X_{DWI}^{(c)}$ denotes the corresponding central T2WI and DWI slice.

By jointly processing MRI intensities, anatomical zonal priors, and lesion localization, the classifier learns anatomy-aware feature representations that exploit both lesion appearance and its anatomical context for csPCa detection. The resulting tensor is processed by a lightweight CNN composed of five convolutional blocks followed by global average pooling and a two-layer multilayer perceptron. Each convolutional block consists of a $3\times3$ convolution, batch normalization, ReLU activation, and max-pooling. The classification network predicts a slice-level csPCa probability, and the maximum probability across all slices from the same patient is used as the patient-level score.

\section{EXPERIMENTS}
\label{experiments}

\noindent\textbf{Datasets} The proposed framework was evaluated on the PI-CAI Public Training and Development dataset~\cite{saha2024artificial}, comprising approximately 1,500 bpMRI studies with T2WI, DWI, and ADC images. Expert csPCa lesion annotations are available for 450 patients, while AI-generated prostate zonal masks are provided for all cases. Five-fold cross-validation was performed using the predefined splits, with results averaged across folds. 

Zonal segmentation generalization was additionally evaluated on an independent cohort of 450 de-identified mpMRI studies collected at Northwestern University under HIPAA compliance was used to evaluate the generalization of the zonal segmentation model. The mpMRI protocol included T2WI, T1WI, DWI, ADC, and dynamic contrast-enhanced imaging, with nearly all scans acquired on 3T scanners. WG, PZ, and TZ annotations were manually delineated on T2WI by two expert radiologists using ITK-SNAP~\cite{py06nimg}.

\noindent\textbf{Implementation Details}
ZH-Mamba was trained in three stages (Fig.~\ref{fig:model}): H-Mamba UNet-S was first trained for lesion localization using three consecutive T2WI and DWI slices, followed by H-Mamba UNet-L for zonal segmentation using three T2WI slices. The zonal network was then frozen while the lesion network and CNN classifier were jointly optimized. Inputs were resized to $256\times256$ and center-cropped to $128\times128$. Models were trained using AdamW ($10^{-3}$), with Dice+BCE loss for segmentation and $\pm15^\circ$ rotation augmentation, on an NVIDIA RTX A6000 GPU. Segmentation was evaluated using mean Dice $\pm$ standard deviation. Patient-level csPCa scores were obtained by maximum aggregation of slice-level probabilities and evaluated using F1-score, AUROC, and $\mathrm{AP}_{\mathrm{Patient}}$. 

\noindent\textbf{Results}
Table~\ref{tab:dice_results} compares zonal segmentation performance on the PI-CAI dataset and an independent internal dataset with representative CNN-, Mamba-, and foundation model-based approaches. Among the compared methods, SAM-Med3D~\cite{wang2025sam} is a prompt-based 3D foundation model, while CAT-Net~\cite{hung2022cat}, a transformer-enhanced nnUNet~\cite{isensee2021nnu}, exploits inter-slice contextual information through a cross-slice attention mechanism despite using a 2D backbone. The proposed 2.5D H-Mamba UNet-L consistently achieves the best performance across all prostate zones, reaching Dice scores of 0.95/0.85/0.95 (WG/PZ/TZ) on PI-CAI and 0.93/0.80/0.91 on the internal dataset. These results demonstrate that the proposed Hadamard-based feature fusion improves segmentation accuracy while maintaining a lightweight architecture that requires only three consecutive T2WI slices.

Table~\ref{tab:lesion_results} summarizes lesion localization and patient-level csPCa detection performance. For lesion segmentation, H-Mamba UNet achieves strong Dice performance, outperforming representative CNN-, Transformer-, and Mamba-based methods. Furthermore, it outperforms prompt-based SAM models, which are evaluated only for lesion segmentation because they require user-provided prompts. The accurate lesion localization produced by H-Mamba provides reliable lesion priors for the downstream classification stage. Combined with the propagated anatomical zonal information from the segmentation network, the proposed ZH-Mamba framework achieves an AUC of 0.86, an AP of 0.72, and an F1-score of 0.76, demonstrating that explicit propagation of anatomical zonal priors improves patient-level csPCa detection beyond lesion localization alone.

\begin{figure}[ht]
\centering
\includegraphics[width=0.8\textwidth, height=5cm]{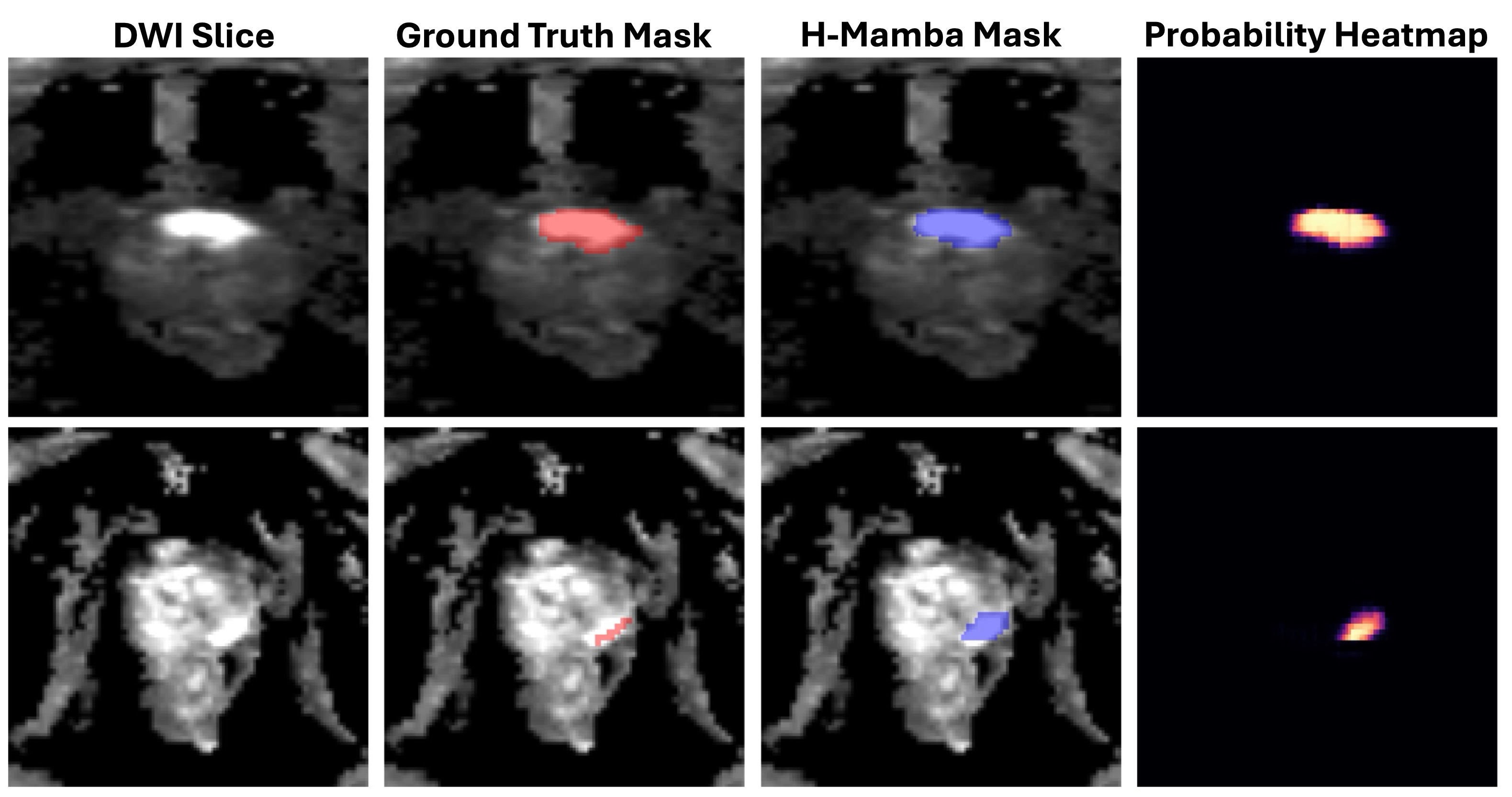}
\caption{Qualitative lesion localization results. Representative DWI slices with ground-truth lesion masks, H-Mamba predictions, and corresponding probability heatmaps.}
\label{fig:prediction}
\end{figure}

Figure~\ref{fig:prediction} provides qualitative examples of lesion localization using the proposed H-Mamba UNet. The predicted masks closely correspond to the ground-truth lesion regions, while the probability heatmaps show concentrated responses around the suspected lesions. These examples illustrate that the H-Mamba architecture preserves spatially localized lesion information while incorporating the proposed transform-domain feature fusion.

\begin{table} 
\centering
\caption{H-Mamba zonal segmentation performance (Dice) on the PI-CAI dataset and the independent internal dataset. Values are reported as mean $\pm$ standard deviation across the 5-fold cross-validation, except for the WG Dice scores, which is omitted for space limitations. Param denotes the number of learnable parameters in Millions (M).}
\label{tab:dice_results}

\begin{tabular}{|l|ccc|ccc|c|}
\hline
 & \multicolumn{3}{c|}{PI-CAI} & \multicolumn{3}{c|}{Internal Dataset}& \multicolumn{1}{c|}{Param}   \\
Model & WG & PZ & TZ & WG & PZ & TZ & M\\
\hline
SAM-Med3D~\cite{wang2025sam}
&0.85 &0.55$\pm$0.05 & 0.80$\pm$0.03 
& 0.81& 0.52$\pm$0.06& 0.79$\pm$0.06& 100.0 \\
CAT-Net \cite{hung2022cat} 
& 0.89 & 0.83$\pm$0.04 & 0.90$\pm$0.04 
& 0.85 & 0.78$\pm$0.05 & 0.85$\pm$0.05 & 40.0 \\
Mamba UNet \cite{wang2024mamba} 
& 0.93 & 0.80$\pm$0.05 & 0.93$\pm$0.05
& 0.92 & 0.79$\pm$0.06 & 0.90$\pm$0.05 & 33.2\\
\hline
H-Mamba UNet-S
& 0.93 & 0.79$\pm$0.05 & 0.93$\pm$0.04 
& 0.91 & 0.75$\pm$0.05 & 0.90$\pm$0.06 &\textbf{8.9} \\

H-Mamba UNet-L
& \textbf{0.95} & \textbf{0.85$\pm$0.04} & \textbf{0.95$\pm$0.03}
& \textbf{0.93} & \textbf{0.80$\pm$0.05} & \textbf{0.91$\pm$0.05} & 33.0 \\

\hline
\end{tabular}

\end{table}

\begin{table}
\centering
\caption{ZH-Mamba performance comparison in csPCa localization and detection on PI-CAI dataset. Values are reported as mean $\pm$ standard deviation across the 5-fold cross-validation. Param denotes the number of learnable parameters in Millions (M).}
\label{tab:lesion_results}

\begin{tabular}{|l|c|c|c|c|c|c|}
\hline
Model & Param (M) & Dice  & F1 & AUC & AP  \\
\hline
MedSAM2~\cite{ma2025medsam2}&38.9 &0.57$\pm$0.05 & - & -& -\\
Attention UNet 2D \cite{oktay2018attention} 
&30.0& 0.56$\pm$0.05  &0.60  & 0.61 & 0.50  \\
nnU-Netv2~\cite{isensee2024nnu}&31.1 & 0.57$\pm$0.04 & 0.62 & 0.82 & 0.61\\
Mamba UNet 2D \cite{wang2024mamba}&33.2& 0.62$\pm$0.05  & 0.66  & 0.79 & 0.60 \\
\hline
\textbf{ZH-Mamba UNet-S 2.5D (ours)}& \textbf{10.5}  & 0.69$\pm$0.04 & 0.74 & \textbf{0.86} & 0.70  \\
\textbf{ZH-Mamba UNet-L 2.5D (ours)}& 34.7  & \textbf{0.70$\pm$0.04} & \textbf{0.76} &\textbf{0.86} &\textbf{0.72}  \\
\hline
\end{tabular}

\end{table}

\begin{table}[t]
\centering
\caption{Component-wise ablation study of the proposed ZH-Mamba framework on patient-level csPCa detection.}
\label{tab:ablation}

\begin{tabular}{|l|c|c|c|c|c|}
\hline
Configuration &
Zonal &
Lesion &
Hadamard &
AUC &
AP \\
\hline

MRI Only
& \ding{55} & \ding{55} & \ding{55}   & 0.79 & 0.60\\

+ Lesion Localization 
& \ding{55} & \ding{51} & \ding{55}  & 0.80 &0.63\\

+ Zonal Awareness
& \ding{51} & \ding{55} & \ding{55} & 0.81 &0.65\\

+ Zonal Awareness + Lesion Localization
& \ding{51} & \ding{51} & \ding{55} & 0.82 &0.66\\
\textbf{+Hadamard Fusion (ZH-Mamba)}
& \ding{51}& \ding{51} & \ding{51}  & \textbf{0.86} & \textbf{0.72}  \\
\hline

\end{tabular}
\end{table}

\noindent\textbf{Ablation Study.}
Table~\ref{tab:ablation} evaluates the effect of the number of input slices and the contribution of anatomical and lesion priors to ZH-Mamba. Increasing the input from a single slice to three consecutive slices improves segmentation performance by providing inter-slice contextual information, while further increasing the number of slices provides limited additional benefit, motivating the use of three slices in the proposed framework. For csPCa detection, progressively incorporating the predicted lesion and zonal masks improves patient-level performance compared with MRI alone, demonstrating that lesion localization and prostate zonal anatomy provide complementary information for classification. While the zonal segmentation network is independently evaluated on our internal dataset, future work will extend this dataset with csPCa annotations to externally validate the complete ZH-Mamba framework.

\section{Conclusion}

In this work, we proposed ZH-Mamba, a zonal-aware framework integrating prostate zonal segmentation, lesion localization, and clinically significant prostate cancer (csPCa) classification within a unified pipeline. H-Mamba incorporates a learnable Hadamard fusion mechanism to efficiently combine spatial- and transform-domain representations with minimal additional parameters. Experimental results demonstrate accurate zonal segmentation, lesion localization, and strong patient-level csPCa detection on the PI-CAI dataset. By propagating anatomical zonal information throughout the diagnostic pipeline using only a small number of MRI slices, ZH-Mamba demonstrates the potential of zonal-aware modeling for automated prostate cancer diagnosis.

\section{Acknowledgments}
This work was supported in part by the National Institutes of Health
(NIH) under Grant U01-CA268808 and the National Science Foundation
(NSF) IDEAL program.

\bibliography{report} 
\bibliographystyle{spiebib} 

\end{document}